\documentclass[12pt,preprint]{aastex}

\def\apj{\emph {ApJ}}

\def\apjl{\emph {ApJ Lett.}}
\def\mnras{\emph {MNRAS}}
\def\aap{\emph {A\&A}}
\def\aaps{\emph {A\&A Supp.}}
\def\araa{\emph {ARAA}}
\def\prl{\emph {PRL}}

\begin{document}
\title{Microwave Emission from Aligned Dust}
 \author{A. Lazarian}
\affil{Department of Astronomy, University of Wisconsin, Madison, WI 53706}
 \author{D. Finkbeiner}
\affil{Department of Astrophysical Sciences, Princeton University, 
Princeton, NJ 08544}

\begin{abstract}
Polarized microwave emission from dust is an important foreground that
may contaminate polarized CMB studies unless carefully accounted
for. We discuss potential difficulties associated with this
foreground, namely, the existence of different grain populations with
very different emission/polarization properties and variations of the
polarization yield with grain temperature. In particular, we discuss
observational evidence in favor of rotational emission from tiny PAH
particles with dipole moments, i.e. ``spinning dust'',
and also consider magneto-dipole emission from strongly magnetized
grains. We argue that in terms of polarization, the magneto-dipole emission
may dominate even if its contribution to total emissivity is subdominant.
Addressing polarized emission at frequencies
larger than $\sim100$ GHz, we discuss the complications
arising from the existence of dust components with different temperatures
and possibly different alignment properties. 

\end{abstract}

\maketitle

\section{Introduction}

Diffuse Galactic microwave emission carries important information on
the fundamental
properties of the interstellar medium, but it also interferes with
Cosmic Microwave Background (CMB) experiments
(see Bouchet et al. 1999, Tegmark et al. 2000,
Efstathiou 2003, this volume ). 
Polarization of the CMB provides information about the Universe
that is not contained in the temperature data alone. In particular, it
offers a unique
way to specifically trace the primordial perturbations of tensorial nature
({\em i.e.} cosmological gravitational waves, see Seljak \& Zaldarriaga 1997, 
Kamionkowski et al. 1997, Kamionkowski 2003, this volume),
and allows one to break some important degeneracies that remain in the 
measurement of cosmological
parameters with intensity alone (Zaldarriaga et al. 1997, Davis 
\& Wilkinson 1999, 
Lesgourgues et al. 1999, Prunet et al. 2000).
Therefore, a number of groups
around the world (see Table~1 in Staggs et al. 1999) work hard to measure
the CMB polarization. The first exciting measurements of CMB
polarization have recently been reported
(Carlstrom 2003, this volume, and Page et al. 2003).  The 
polarization of Galactic emission, long of interest for ISM studies, is
now also an important foreground for cosmology.

Among different sources of polarized foregrounds, interstellar dust is probably
the most difficult to deal with, for many reasons.
First of all, dust has both a population of tiny grains (Leger \&
Puget 1984), which are frequently called PAH, in addition to the ``classical''
power-law distribution of larger grains (Mathis, Rumpl \& Nordsieck 1977).
Then the typical composition of grains changes with their size,
with equilibrium temperature depending on both size and composition
(Draine \& Lee 1984, Finkbeiner et al. 1999).  
The degree of grain alignment may depend on size and composition,
leading to a frequency dependence of the polarization
(Hildebrand et al. 2001). Moreover,
both recent experience with microwave emissivity and theoretical studies
of expected polarization response (Draine \& Lazarian 1999)
show that the naive extrapolation of the grain properties from FIR to
microwave does not work. In addition, in spite of
the evident progress achieved by the grain alignment theory 
(see review by Lazarian 2003), unanswered questions still remain there.

The discovery of the anomalous emission in the range of 10-100~GHz
illustrates well the treacherous nature of dust.
Until very recently it has been thought that
there are three major components of the diffuse microwave Galactic foreground:
synchrotron emission, free-free radiation from plasma (thermal bremsstrahlung)
and thermal emission from dust. In the microwave range of 10-90~GHz the latter
is subdominant, leaving essentially two
components. However, it is exactly in this range that an anomalous
emission was reported (Kogut et al. 1996a, 1996b). In the
paper by de Oliveira-Costa et al. (2002) this emission was nicknamed
``Foreground X'', which properly reflects its mysterious nature.
This component is spatially correlated with 100 $\mu$m thermal
dust emission, but its intensity is much higher than one would expect
by directly extrapolating the thermal dust emission spectrum
to the microwave range.  Similar surprises may await in the 
foreground polarimetry data. 

In this review, we briefly summarize what is known about the grain populations,
grain emission and grain alignment. We discuss the origin of the Foreground
X and its expected polarization. Recent reviews of the subject include
Draine \& Lazarian 1999, Lazarian \& Prunet 2002.

\section{Observational Evidence}

\subsection{Infrared emission: extrapolation to microwave range}

The emission spectrum of diffuse interstellar dust was mostly obtained
by the {\it InfraRed Astronomy Satellite} (IRAS) and infrared
spectrometers on the {\it COsmic Background Explorer} (COBE) and on
the {\it InfraRed Telescope in Space} (IRTS).

The emission at short wavelength, e.g. $<50$~$\mu$m,
arises from transiently heated very small grains. These grains have
such a small heat capacity that the absorption of a single 6 eV starlight
photon raises their temperature to $T>200$K. Typically these grains
have less than $300$ atoms and can be viewed as large molecules
rather than dust particles. They are, however, sufficiently numerous
to account for $\sim 35\%$ of the total starlight absorption.
The thermal (vibrational) emissivity of these grains is thought to be
negligible at low frequency, because they spend most of their time
cold, but emit most of their energy when they are hot. 

The dominant dust emission above $\sim100$ GHz is emission from grains
large enough to be in equilibrium with the interstellar radiation
field.
Emission from this dust peaks at $\sim 140\micron$
and deviates strongly from a thermal blackbody spectrum.  A
Rayleigh-Jeans emissivity function of $\nu^2$ has often been assumed
in the literature (e.g. Draine \& Lee 1984, Schlegel, Finkbeiner \&
Davis 1998) but when dust temperature variation is accounted for, the
COBE FIRAS data (Fixsen et al. 1997) are better fit by a steeper power
law emissivity ($\beta=2.6$) near the peak and $\beta=1.7$ at lower
frequencies, with a break at about $500$ GHz (Finkbeiner et al. 1999).
This fit tied the IRAS and DIRBE data to FIRAS via a fit with only 4
global parameters describing the two emissivity laws, and the
requirement that the emission is dominated by grains in equilibrium
with the interstellar radiation field.  Predictions at $6'$ resolution
based on this fit are available on the web.\footnote{http://skymaps.info}

The two-component model is a substantially better fit (reduced
$\chi^2=1.85$ compared to 31 for a $\nu^2$ model) even when the
spatial and spectral covariance of the FIRAS data (Fixsen
et al. 1997) are included (Finkbeiner et al. 1999).  And the model is
physically plausible: amorphous
silicates with a wide range of emissivity indices $\beta\sim 1.2-2.7$
have been observed in the lab (Agladze et al. 1996), including
amorphous MgO$\cdot$2SiO$_2$ which has a very high microwave
emissivity to optical absorption ratio, leading to rather different
mean temperatures (9K and 16K) for the two components.  These lab
emissivities were measured at $\sim300$ GHz and 20K and may become
steeper at lower frequencies.  However, this interpretation of the
spectral break is hardly unique; if the dominant emitter has such a
break in its emissivity function at 500 GHz, then a single component
could explain the data just as well.  Another explanation that has
been advanced is very cold dust grains spatially mixed with the warm
dust (Reach et al. 1995), though a physical mechanism
for keeping the grains so cold is not proposed.  Such a
model would presumably predict a steeper slope at lower frequencies as
well.

Regardless of interpretation, the Finkbeiner et al. (1999) model
has been very successful in the sub-mm - microwave, though small but
interesting deviations from the model have been observed by BOOMERANG
(Masi et al. 2001).  At lower frequencies, however, there is a
surprise. 

Comparing these predictions to COBE DMR, Finkbeiner et al. found that
COBE 90 GHz was slightly higher, but at 53 and 31 GHz the emission per
dust column is a factor of 2.2 and 31 higher than expected.  These
results are similar to the earlier Kogut et al. (1996) results derived without
an explicit dust temperature correction.  Because of this it was
expected that the FIRAS-based predictions would agree well with WMAP
94 GHz, but be significantly contaminated by some other dust-correlated
emission mechanism at lower frequencies, and this appears to be true. 
Until this other emission is understood, extrapolation of far IR
polarization measurements to the microwave regime will be perilous. 

\subsection{Anomalous microwave emission}

The first detection of anomalous dust correlated emission by COBE
(Kogut et al. 1996a, 1996b) was quickly followed by detections in the data
sets from Saskatoon (de Oliveira-Costa et al. 1997), OVRO (Leitch et
al. 1997), the 19~GHz survey (de Oliveira-Costa et al. 1998), and
Tenerife (de Oliveira-Costa et al. 1999, Mukherjee et al. 2000).
Initially, the anomalous emission was identified as thermal  bremsstrahlung
from ionized gas correlated with dust (Kogut et al. 1996a) and presumably
produced
by photoionized cloud rims (McCullough et al. 1999). This idea was
scrutinized in Draine \& Lazarian (1997) and criticized on
energetic grounds. Poor correlation of H$\alpha$ with 100~$\mu$m
emission also argued against the free-free explanation (McCullough et
al. 1999). These arguments are summarized in Draine
\& Lazarian (1999). Recently
de Oliveira-Costa et al. (2000) used Wisconsin H-Alpha Mapper (WHAM)
survey data and established that the free-free emission ``is about an order
of magnitude below Foreground X over the entire range of frequencies
and latitudes where it is detected''. The authors conclude that the
Foreground X cannot be explained as free-free emission. Additional
evidence supporting this conclusion has come from
a study at 5, 8 and 10~GHz by Finkbeiner et al. (2002) of several dark
clouds and HII regions, two of which show a significantly rising
spectrum from 5 to 10 GHz.

The recent Wilkinson Microwave Anisotropy Probe (WMAP) data were used
to claim a lower limit of 5\% for the spinning dust fraction at 23 GHz
(Bennett et al. 2003).  However, other models of spinning dust are not
ruled out by the WMAP data, and in fact fit reasonably well.
Finkbeiner (2003) performed a fit to WMAP using only a CMB template, a
free-free template (based on H$\alpha$ correlated emission plus hot
gas emission near the Galactic center), a soft synchrotron template
traced by the 408 MHz map, a thermal dust extrapolation (Finkbeiner et
al. 1999) and a spinning dust template consisting of dust column
density times $T^3_{dust}$.  This fit results in excellent
$\chi^2/dof$ values (1.6,1.09,1.08,1.05,1.08) at (23,33,41,61,94) GHz
and a reasonable spectral shape for the average spinning dust
spectrum.  The whole sky $|b| < 30$ degrees was used where H$\alpha$
extinction is less than 2 mag, except for point sources, Orion, and
NGC5090.  The derived emissivities, expressed as Jy/sr$^{-1}$ per H
atom for comparison with DL98b, are shown in Fig. 1.  Note that
there is considerable variation around the sky in the spinning dust
spectrum, and Figure 1. shows only the average.  The data points
\emph{red filled circles} fall somewhat lower than the WNM, CNM, and
WIM ISM models and appear flatter, but some superposition of spinning
dust models would produce this average spectrum. 

\begin{figure}
\resizebox{\textwidth}{15cm}{\includegraphics{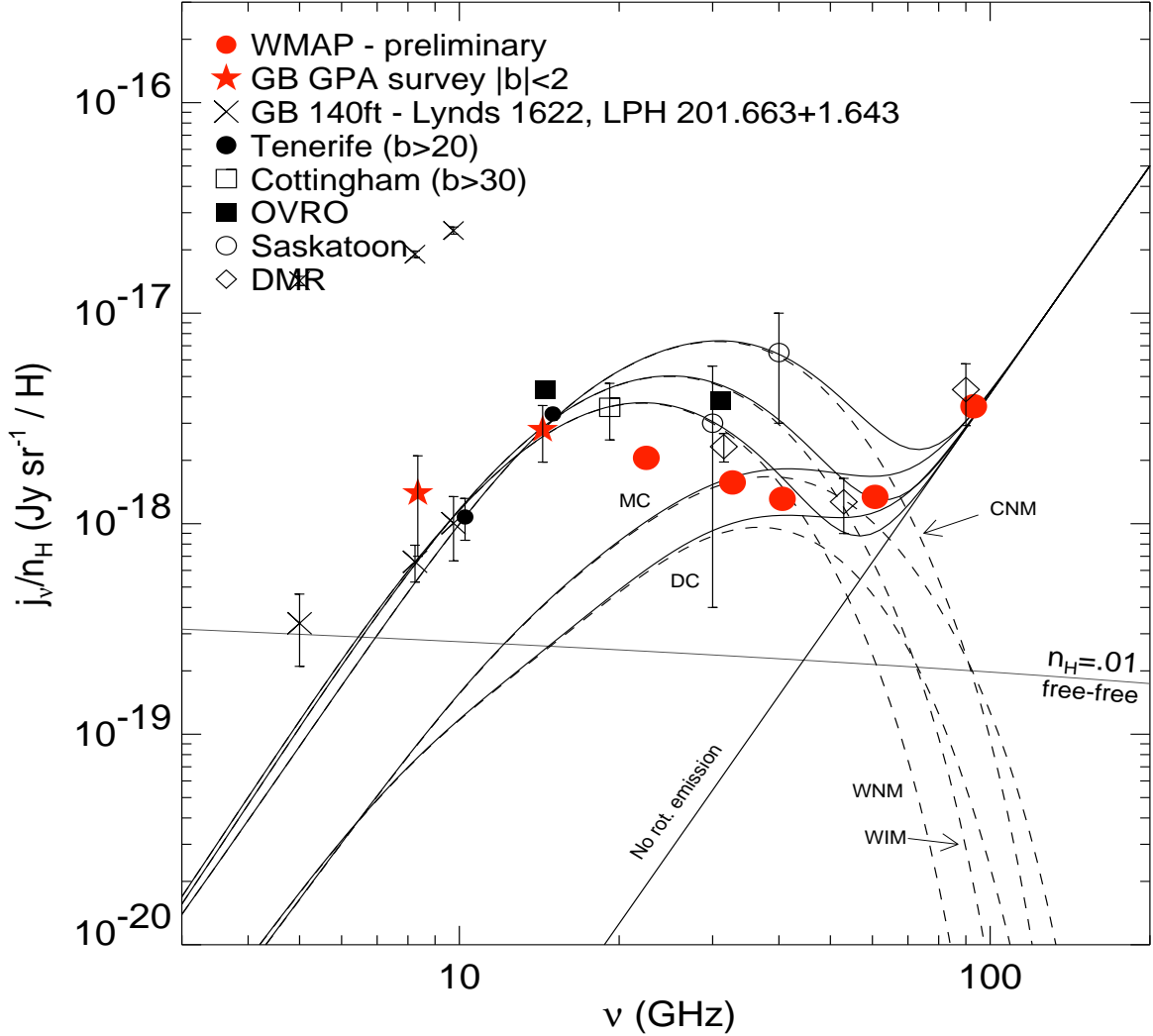}}
\caption{Model dust emissivity per H atom for DC, MC, CNM, WNM, and
WIM conditions (as in Draine \& Lazarian 1998b, Figure 9)
with (\emph{solid lines}) and without (\emph{dashed lines})
contribution from vibrational dust at mean temperature. 
Gray line is emission from free-free for given $n_H$, or rather 
$<n_en_p>/<n_H>$ averaged along the line of sight. 
Also shown are measurements from the COBE/DMR (\emph{open diamonds})
from Finkbeiner et al. (1999), similar to Kogut
et al. (1996); Saskatoon (\emph{open circles}) (de Oliveira-Costa 
et al. 1997); the Cottingham \& Boughn $19.2$ GHz survey 
(\emph{open square}) (de
Oliveira-Costa et al. 1998), OVRO data (\emph{solid squares}) 
(Leitch et al. 1997);
Tenerife data (\emph{solid circles})(de Oliveira-Costa et al. 1999);
GB 140 foot (\emph{crosses})(Finkbeiner et al. 2002), GPA
(\emph{red stars})(Finkbeiner et al. 2003), and WMAP (\emph{red
circles})(Finkbeiner 2003). 
The OVRO points have been lowered a factor of 3 relative to Draine \&
Lazarian (1998b, Figure 9), because the unusual dust temperature near
the NCP caused an underestimate of the H column density along those lines
of sight.  In fact, the H columns used in this plot are actually
derived from SFD $E(B-V)$ with a conversion factor of $8\times10^{21}$
H / mag. 
Given the large range of model curves, all measurements 
are consistent with some superposition of spinning dust, vibrational
dust, and free-free emission. 
}
\end{figure}

This WMAP analysis alone does not rule out the Bennett et al. (2003)
hypothesis of hard synchrotron emission, but when combined with the
Green Bank Galactic Plane survey data (Langston et al. 2000) at 8 and
14 GHz, spinning dust appears to provide a much better fit than hard
synchrotron (Finkbeiner, Langston, \& Minter 2003).  Some caution is
necessary, because the rising ISM spectrum seen from 8 to 14 GHz is
observed in the Galactic plane (red stars in Fig. 1), while the WMAP
fit is done at higher latitudes; however it is currently a good
working hypothesis that spinning dust emission is a substantial
contribution to ISM emission at $10 < \nu < 50$ GHz. 
Other groups analysing WMAP data found more evidence in favor of
spinning dust (Hildebrand, private communication, Lagache 2003).
For instance, recent results in
Hildebrand \& Kirby (2003, preprint) on L1622 (one of the clouds observed
by Finkbeiner et al. (2002) at 5, 8 and 10 GHz) obtained using WMAP
data show a smooth continuation of the spectrum in agreement with the
spinning dust model expectations.

\subsection{Alignment of Classical Dust}

Polarization due to interstellar dust alignment was discovered in the middle of
the last century (Hiltner 1949, Hall 1949) and was studied initially via
starlight extinction and more recently through emission. Correlation of
the polarization with the interstellar magnetic field revealed that electric
vector of light polarized via starlight extinction tend to be parallel
to magnetic field\footnote{The polarizations in emission and in extinction
are orthogonal if they are produced by the same grains.}. This corresponds to grains being aligned with their
longer axes perpendicular to the local magnetic field. Due to the
presence of the stochastic magnetic field, the polarization patterns
are pretty involved.

The existing
data presents a complex picture. It is generally accepted that the observations
indicate that the 
ability to produce polarized light depends on grain size and
grain composition. For instance, a limited UV polarimetry dataset available
indicates that graphite grains tend not to be aligned 
(see Clayton et al. 1997),
while maximum entropy technique applied to the existing data
by Martin \& Kim (1995) show that large $>6\times 10^{-6}$~cm
grains are responsible for the polarization via extinction.

Moreover, the environment of grains seems to matter a lot (Goodman 1995,
Lazarian, Goodman \& Myers 1997). A study by Arce et al. (1998) indicates
that grains selectively extinct starlight up to optical depth $A_v<3$.
Recent emission studies (Hildebrand et al. 1999, 2001) produced a polarization spectrum
for dense clouds that reveal a tight correlation between grain temperature
and its ability to emit polarized light. As multicomponent fits
invoking grains of different temperature
were claimed to provide a better fit for the observed 1~mm-100~$\mu$m
emission (see Finkbeiner, Schlegel \& Davis 1999), this correlation may be 
very troublesome
for the attempts to construct polarization templates. 

The balloon-borne Archeops mission detects polarization at 353 GHz
(850$\mu$m) at the level of 4-5\%, and over 10\% in some clouds
(Benoit et al. 2003).  This is about the level expected based on
polarization of starlight and emission at shorter wavelengths.  We
eagerly await polarization data from WMAP at $23-94$ GHz.

\section{Polarized Emission from Classical Dust}

The basic explanation of polarized radiation from dust is straightforward.
Aligned dust particles preferentially extinct (i.e. absorb and scatter)
the $E$-component of starlight parallel to their longer axis.
The $E$-component of the emitted thermal radiation, on the contrary,
is higher along the
longer axis. Thus for aligned grains one must have polarization.
What is the cause of alignment?

Grain alignment is an exciting and very rich area of research. For example,
two new solid state effects have been discovered recently in the process
of understanding grain dynamics (Lazarian \& Draine 1999, 2000).
It is known that a number of 
mechanisms can provide grain alignment (see review by Lazarian
2000 and Table~1 in Lazarian, Goodman, \& Myers 1997). Some of
them rely on paramagnetic dissipation of rotational energy 
(Davis-Greenstein 1951, Purcell 1979, Mathis 1986, Lazarian \& Draine 1997,
Lazarian 1997a, Roberge \& Lazarian 1999) , some
appeal to the anisotropic gaseous bombardment when a grain moves
supersonically through the ambient gas (Gold 1951, Purcell \& Spitzer 
1971, Dolginov \& Mytrophanov 1976, Lazarian 1994, 1997b, Roberge,
Hanany \& Messinger 1995, Lazarian
\& Efroimsky 1996). Many grains are definitely paramagnetic and some may be
strongly magnetic. Supersonic grain motions may be due to
outflows (Purcell 1969),
MHD turbulence (Lazarian 1994, Lazarian \& Yan 2002, Yan \& Lazarian
2003) or ambipolar diffusion (Roberge
\& Hanany 1990).

At present,
grain alignment
via radiative torques (Draine \& Weingartner 1996, 1997)
looks preferable, although the theory and the understanding of the mechanism
are far from being complete (see review by Lazarian 2003). 
The mechanism appeals to a spin-up of a grain
as it differentially scatters left and right polarized photons (
Dolginov 1972, Dolginov \& Mytrophanov 1976). 
This process acts efficiently if the irregular
grain has its size comparable with the photon wavelength. The mechanism
can account for the systematic variations
of the alignment efficiency with extinction. 
However, other mechanisms should
also work. For instance, the
paramagnetic mechanism may preferentially act on small grains (Lazarian
\& Martin 2003, in preparation), while
mechanical alignment may act in the regions of outflows (Rao
et al. 1998). In general, the variety of astrophysical conditions allows
various mechanisms to have their niche. 

Note that in interstellar environments grain alignment
respects the magnetic field orientation, even if the mechanism of alignment
is not magnetic in nature.  This is because
the Larmor precession of grains is so fast compared to the time scales
over which either the magnetic field changes its direction or the alignment
mechanism acts. In general, the alignment 
may happen either parallel and perpendicular to the magnetic field. 
However, in most cases, the alignment happens with long grain axes
perpendicular to magnetic field.

Alignment of grains is different in diffuse gas and molecular clouds.
Lazarian, Myers \& Goodman (1997) showed that in dark clouds
without star formation all alignment mechanisms fail. Indeed,
grain alignment depends on non-equilibrium processes while interiors
of dark clouds are close to thermodynamic equilibrium.
As soon as stars are born within clouds, the conditions in their vicinity
become favorable for grain alignment. 
This explains why far infrared
polarimetry detects aligned grains, while near infrared and optical polarimetry
does not. 

We may hope that grain alignment in diffuse clouds is more uniform.
Radiation freely penetrates them and therefore the radiative torques
must ensure good alignment.
This assumption was used by Fosalba et al (2001)
to relate the polarization from dust extinction and the polarization
from dust emission. Discussion of this problem is presented in
Cho \& Lazarian (2003, this volume).

As we have already discussed, grain alignment traces the direction of
the local magnetic field.
In the presence of turbulence, this field is very complex. It was
shown by Cho \& Lazarian (2002a) that MHD turbulence can explain
the spatial variations of both synchrotron emission and starlight
polarization. We note when we deal with dust aligned
with a turbulent magnetic field, the resulting
polarization depends on the telescope resolution at a particular
wavelength. A possible way of dealing with this complication is to
correct for the field stochasticity. A tensor description 
of turbulent magnetic fields was 
obtained in Cho, Lazarian \& Vishniac (2002) and this can be used for this
purpose (see also Cho \& Lazarian 2002b).   The corresponding
 research should also
yield insight into the operation of the Galactic dynamo, high latitude
MHD turbulence, and turbulent mixing, and will lead to many yet
unforeseen discoveries.

\section{Polarized Emission from Spinning Dust}

Can the ultrasmall grains observed via Mid-IR be important at the microwave
range? The naive answer to this question is no, as the total mass in those
grains is small.
However, DL98a considered a different mechanism of emission,
namely, the rotational emission\footnote{The very idea of grain
rotational emission was first
discussed by Erickson (1957). More recently,
after the discovery of the population of ultrasmall grains,
Ferrara \& Dettmar (1994) noted that the rotational emission from such
grains may be observable, but their treatment assumed Brownian
thermal rotation of grains, which is not true.}
that must emerge when a grain with a dipole moment $\mu$  rotates
with angular velocity\footnote{The calculations in DL98a were questioned
by Ragot (2002) who considered the effect of plasma wave drag on spinning
dust grains. However, the treatment of ionized particles as a continuous
plasma when less than a few particles have chance to interact with the
grain over its period does not seem to be right. Moreover, it is possible
to show that if it were right, the plasma would not be transparent to
microwave emission.} $\omega$.

For the model with the most
likely set of parameters, DL98a obtained a reasonable fit with observations
available at that time. It is extremely important that new data points
obtained later (de Oliveira-Costa et al. 1998,
de Oliveira-Costa et al. 1999) correspond to the already published
model. The observed flattening of the spectrum and its turnover
around $20$~GHz agree well with the spinning dust predictions.

Microwave emission from spinning grains is expected to be polarized if
grains are aligned. Alignment of ultrasmall grains 
(essentially large molecules) is likely to be different from alignment
of large (i.e. $a>10^{-6}$~cm) grains.
One of the mechanisms that might produce alignment of the ultrasmall
grains is the
paramagnetic dissipation mechanism
of Davis and Greenstein (1951). The Davis-Greenstein alignment mechanism is
straightforward: for a spinning grain
the component of interstellar magnetic field
perpendicular to the grain angular velocity varies in grain coordinates,
resulting in time-dependent magnetization, associated
energy dissipation, and a torque acting on the grain.
As a result grains  tend to rotate
with angular momenta parallel to the
interstellar magnetic field.

Lazarian \& Draine (2000, henceforth LD00) found
that the traditional picture
of paramagnetic relaxation is
incomplete, since it
disregards the so-called ``Barnett magnetization'' (Landau \& Lifshitz 1960).
The Barnett effect, the inverse of the Einstein-De Haas effect,
consists of the spontaneous magnetization of
a paramagnetic body
rotating in field-free space. This effect can be understood in
terms of the lattice sharing part of its angular momentum with
the spin system. Therefore the implicit assumption in Davis
\& Greenstein (1951)--
that the magnetization within a {\it rotating grain} in a {\it static}
magnetic field is equivalent to the magnetization within a
{\it stationary grain} in a {\it rotating} magnetic field --
is clearly not exact.

LD00 accounted for the ``Barnett magnetization'' and termed the effect
of enhanced relaxation arising from grain magnetization ``resonance
relaxation''. It is clear from Fig.~2 that resonance relaxation persists
at the frequencies when the Davis-Greenstein relaxation vanishes. However
the polarization is marginal for $\nu>35$~GHz anyhow. The discontinuity
at $\sim 20$~GHz
is due to the assumption that smaller grains are planar, and larger
grains are spherical. The microwave emission will be polarized
in the plane perpendicular to magnetic field.

\begin{figure}
\resizebox{\textwidth}{12cm}{\includegraphics{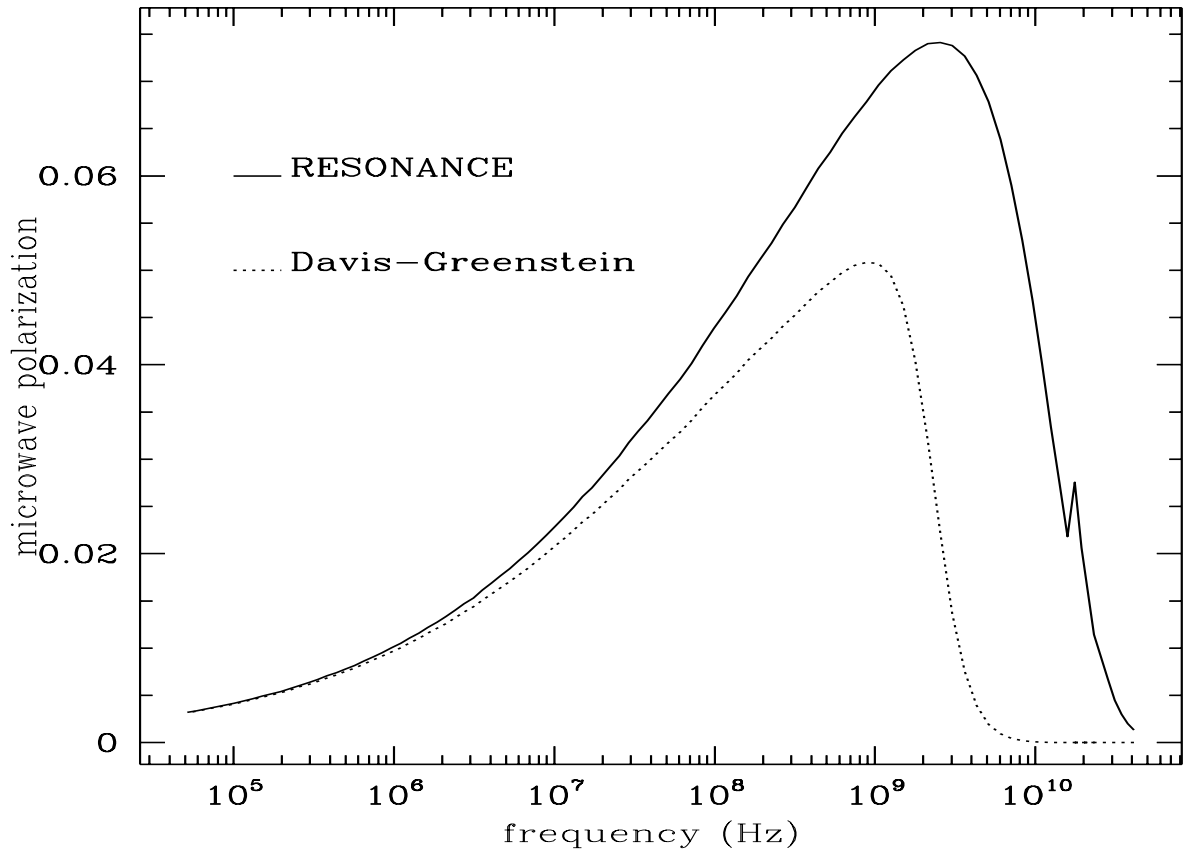}}
\caption{Polarization for both
        resonance relaxation and Davis-Greenstein relaxation for grains in
        the cold interstellar medium as a function of frequency (from LD00).
For resonance relaxation the saturation effects
(see eq.\ (1)) are neglected, which means
that the upper curves correspond to the {\it maximal} values allowed by the
paramagnetic mechanism.}
\end{figure}

Can we check the alignment of ultrasmall grains via infrared polarimetry?
The answer to this
question is ``probably not''. Indeed, as discussed earlier,
infrared emission from ultrasmall grains, e.g. 12 $\mu$m emission,
takes place as grains absorb UV photons. These photons raise
grain temperature, randomizing grain axes in relation to
its angular momentum (see Lazarian \& Roberge 1997). Taking values
for Barnett relaxation from Lazarian \& Draine (1999), we get
the randomization time of the $10^{-7}$~cm grain to be
 $2\times 10^{-6}$~s, which is less than the grain cooling time. As a
result, the emanating infrared radiation will be polarized very marginally.
If, however, Barnett relaxation is suppressed, the randomization time
is determined
by inelastic relaxation (Lazarian \& Efroimsky 1999) and is
$\sim 0.1$~s, which would entail a partial polarization of
infrared emission.

\section{Polarized Emission from Magnetic Grains}

While the spinning grain hypothesis got recognition in the
community, the magnetic dipole emission model suggested by Draine
\& Lazarian (1999, henceforth DL99) was left essentially unnoticed.
This is unfortunate, as magnetic dipole emission provides a
possible alternative explanation for the Foreground X. Magnetic
dipole emission is negligible at optical and infrared frequencies.
However, when the
frequency of the oscillating magnetic field approaches the precession
frequency of electron spin in the field of its neighbors, i.e.
$10$~GHz, the magneto dipole emissivity becomes substantial.

How likely is that grains are strongly magnetic? Iron is the fifth
most abundant element by mass and it is well known that it resides in
dust grains (see Savage \& Sembach 1996). If $30\%$ of grain mass is
carbonaceous, Fe and Ni contribute approximately $30\%$ of the remaining
grain mass. Magnetic inclusions are widely discussed in grain
alignment literature (Jones \& Spitzer 1967, Mathis 1986, Martin 1995,
Goodman \&
Whittet 1996).
If a substantial part of this material is ferromagnetic
or ferrimagnetic, the magneto-dipole emission can be comparable to that
of spinning grains. Indeed, calculations in DL99 showed that less than
$5\%$ of interstellar Fe in the form of metallic grains or inclusions
is necessary to account for the Foreground X at 90~GHz, while magnetite,
i.e. Fe$_3$O$_4$,
can account for a considerable part of the anomalous emissivity over
the whole range of frequencies from 10 to 90~GHz. Adjusting the
magnetic response of the material, i.e. making it more strongly magnetic
than magnetite, but less magnetic than pure metallic Fe, it is possible
to get a good fit for the Foreground X (DL99).

How can magneto-dipole emission be distinguished from that from
 spinning grains?
The most straightforward way is to study microwave emission from regions
of different density. The population of small grains is
depleted in dark clouds (Leger and Puget 1984)
and this should result in a decrease of
contribution from spinning grains.
Private communication from Dick Crutcher who attempted such measurements
corresponds to this tendency, but the very detection of microwave
emissivity is a 3$\sigma$ result.
Obviously the corresponding  measurements are
highly desirable.
As for now,
magnetic grains remain a strong candidate process for producing
part or even all of Foreground X.
In any case, even if magnetic
grains provide subdominant contribution, this can be important
for particular cases of CMB and interstellar studies. For instance,
polarization from magnetic grains may dominate that from spinning
grains even if the emission from spinning grains is more of higher
level.

The mechanism for producing polarized
magneto-dipole emission is similar to that
producing polarization of electro-dipole  thermal emission
emitted from aligned non-spherical grains (see Hildebrand 1988).
There are two
significant differences, however. First, strongly magnetic
grains can contain just a single magnetic domain. Further magnetization
along the axis of this domain is not possible and therefore the
magnetic permeability of the grains gets anisotropic: $\mu=1$ along
the domain axis, and $\mu=\mu_{\bot}$ for a perpendicular direction.
Second, even if a grain contains tiny magnetic inclusions and can be
characterized by isotropic permeability, polarization that it produces
is orthogonal to the electro-dipole radiation emanating through
electro-dipole vibrational emission. In case
of the electro-dipole emission, the longer grain axis defines the vector
of the electric field, while it defines the vector of the
magnetic field in case
of magneto-dipole emission.

The results of calculations for single domain iron particle (longer axis
coincides with the domain axis) and a grain with metallic Fe inclusions
are shown in Fig.~3. Grains are approximated by ellipsoids $a_1<a_2<a_3$
with ${\bf a_1}$ perfectly aligned
parallel to the interstellar magnetic field ${\bf B}$. The polarization
is taken to be positive when the electric vector of emitted radiation
is perpendicular to ${\bf B}$; the latter is the case for electro-dipole
radiation of aligned grains. This is also true (see Fig.~3) for high
frequency radiation from single dipole grains. It is easy to see
why this happens. For high frequencies $|\mu_{\bot}-1|^2\ll 1$
and grain shape factors are unimportant. The only important thing is
that the magnetic fluctuations happen perpendicular to ${\bf a_1}$.
With ${\bf a_1}$ parallel to ${\bf B}$, the electric fluctuations
tend to be perpendicular to ${\bf B}$ which explains the polarization
of single domain grain being positive. For lower frequencies magnetic
fluctuations tend to happen parallel to the intermediate size axis
${\bf a_2}$. As the grain rotates about ${\bf a_1}\|{\bf B}$,
the intensity in a given direction reaches maximum when an observer
sees the ${\bf a_1} {\bf a_2}$ grain cross section. Applying earlier
arguments it is easy to see that magnetic fluctuations are parallel
to ${\bf a_2}$ and therefore for sufficiently large $a_2/a_1$ ratio
the polarization is negative. {\it
The variation of the polarization direction with
frequency presents the characteristic signature of magneto-dipole emission
from aligned single-dipole grains and it can be used to separate this
component from the CMB signal}. Note that the degree of polarization is
large, and such grains may substantially interfere with attempts
at CMB polarimetry. Even if the
intensity of magneto-dipole emission is subdominant
to that from rotating grains, it can still be quite important in
terms of polarization.
A relatively weak polarization response is expected for grains with
magnetic inclusions (see Fig.~3). The resulting emission is negative
as magnetic fluctuations are stronger along longer grain axes, while
the short axis is aligned with ${\bf B}$.

\begin{figure}
\resizebox{\textwidth}{12cm}{\includegraphics{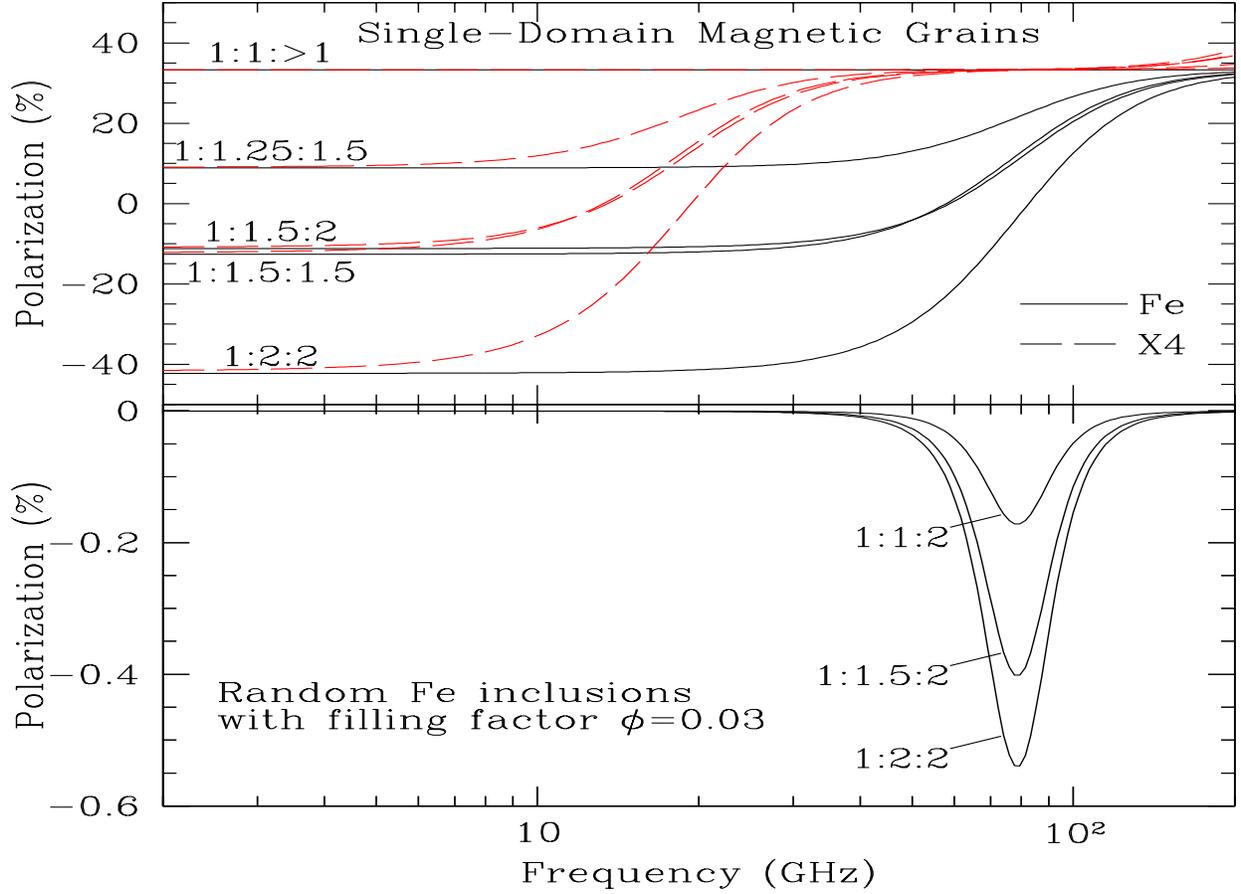}}
\caption{ Polarization from magnetic grains (from DL99). Upper panel:
Polarization of thermal emission from perfectly aligned single
domain grains of metallic Fe (solid lines) or hypothetical magnetic
material that can account for the Foreground X (broken lines).
Lower panel: Polarization from perfectly aligned grains with
Fe inclusions (filling factor is 0.03). Grains are ellipsoidal and
the result are shown for various axial ratios.
}
\end{figure}

Systematic studies of dust foreground polarization should improve
our insight into the formation dust, its structure, its composition.
For instance, DL99 showed that the present-day microwave measurements
do not allow more than 5\% of Fe to be in the form of metallic iron.
More laboratory measurements of microwave properties of candidate
materials are also necessary. Some materials, e.g. iron, were studied
at microwave range only in the 1950's and this sort of data must be checked
again using modern equipment.

\section{Summary}

The principal points discussed above are as follows:

\begin{itemize}

\item Dust provides the most intricate pattern of polarized radiation.
The dependence of polarization of grain temperature, composition,
size and environment makes the use of templates difficult.

\item If anomalous emission in the range of  10-100~GHz is due to spinning
dust particles, the polarization of the emission is marginal for
frequencies larger than $\sim 35$~GHz. If the anomalous emission or
part of it is due to magneto-dipole mechanism the polarization may
be substantial and may exhibit reversals of direction with frequency.

\item To get a better insight into the microwave properties of dust
more laboratory studies are necessary. Some of them, e.g. measurements
of the magnetic susceptibility of candidate materials at microwave 
frequencies, are straightforward using the modern technology.

\end{itemize}

AL was writting this review in a stimulating atmosphere of
 Ecole Normale Superier and
he is happy to acknowledge his ENS Visiting Professor position there. 
DF is a Hubble Fellow.


\begin{thebibliography}{}
\bibitem[]{} Agladze, N. I, Sievers, A. J., Jones, S. A., Burlitch, J. M., \& 
Beckwith, S. V. W. 1996, \apj, 462, 1026
\bibitem[]{}  Altshuler, S.A. \& Kozyrev, B.M. 1964, Electron
Paramagnetic Resonance, Academic Press, New York
\bibitem[]{} Arce, H.G. et al. 1998, \apj, 499L, 93
\bibitem[]{}  Atherton, N.M.\ 1973, Electron Spin Resonance, John Willey \&
Sons, New York
\bibitem[]{} Bennett, C. L., et al. 2003, \apj\ in press, astro-ph/0302208
\bibitem[]{} Benoit, A., et al. 2003, astro-ph/0306222
\bibitem[]{} Bouchet, F.R., Prunet, S., \& Sethi, S.K. \ 1999, \mnras, 302, 663
\bibitem[]{} Boulanger, F., et al. 1996, \aap, 312, 256
\bibitem[]{} Bloch, F. 1946, Phys. Rev., 70, 460
\bibitem[]{} Boulanger, F., \& P\'erault, M. 1988, \apj, 330, 964
\bibitem[]{} Cho, J., Lazarian, A. 2002a, ApJ, 575, 63
\bibitem[]{} Cho, J., Lazarian, A. 2002b, Phys. Rev. Lett., 88, 245001
\bibitem[]{} Cho, J., Lazarian, A. \& Vishniac, E. \ 2002, ApJ, 564, 291
\bibitem[]{} Clayton et al. \ 1997, AJ, 114, 1132
\bibitem[]{} Davis, L., \& Greenstein, J.L. 1951, \apj, 114, 206
\bibitem[]{} Davies, R.D., \& Wilkinson, A. 1999 in ASP Conf. Ser. Vol. 181,
``Microwave Foregrounds'', eds. Angelica de Oliveira-Costa and Max Tegmark,
(San Francisco: ASP), 77 (henceforth ``Microwave Foregrounds'')
\bibitem[]{} Draine, B.T. \& Weingartner, J.C. 1996, ApJ, 470, 551
\bibitem[]{} de Oliveira-Costa, {\it et al.}\ 1997, \apjl, 482, L17
\bibitem[]{} de Oliveira-Costa, {\it et al.}\ 1998, \apjl, 509, L9
\bibitem[]{} de Oliveira-Costa, {\it et al.}\ 1999, \apjl, 527, L9
\bibitem[]{} de Oliveira-Costa, A. et al. 2002, ApJ, 567, 363
\bibitem[]{} D\'esert, Boulanger, F. \& Puget, J.L.\ 1990, A\& A 237, 215
\bibitem[]{} Dolginov A.Z. 1972, Asr. and Space Science, 16, 337
\bibitem[]{} Dolginov A.Z. \& Mytrophanov, I.G. 1976, Asr. and Space Science , 43, 291
\bibitem[]{} Draine, B.T., \& Lazarian, A. 1998a, \apjl, 494, L19 (DL98a)
\bibitem[]{} Draine, B.T., \& Lazarian, A. 1998b, \apj, 508, 157 (DL98b)
\bibitem[]{} Draine, B.T., \& Lazarian, A. 1999, \apj, 512, 000 (DL99)
\bibitem[]{} Draine, B.T., \& Lazarian, A. 1999 in ``Microwave Foregrounds'',
133
\bibitem[]{} Draine, B.T., \& Lee, H.-M. 1984, \apj, 285, 89
\bibitem[]{} Draine, B.T., \& Weingartner, J.C.\ 1996, ApJ, 470, 551
\bibitem[]{} Draine, B.T., \& Weingartner, J.C.\ 1997, ApJ, 480, 633
\bibitem[]{} Gold, T. 1951,  Nature, 169, 322
\bibitem[]{} Erickson, W.C. 1957, \apj, 126, 480
\bibitem[]{} Ferrara, A., \& Dettmar, R.-J. 1994, \apj, 427, 155
\bibitem[]{} Finkbeiner, D. P., Davis, M., \& Schlegel, D. J. 1999, \apj, 524, 867
\bibitem[]{} Finkbeiner, D. P., Schlegel, D. J., Frank, C., \& Heiles, C. 2002,
\apj, 566, 898
\bibitem[]{} Finkbeiner, D. P. 2003, submitted to ApJ
\bibitem[]{} Finkbeiner, D. P., Langston, G., \& Minter, A. 2003, in prep. 

\bibitem[]{} Fosalba, P., Lazarian, A., Prunet, S. \& Tauber, J.A.
\ 2001, ApJ, 
\bibitem[]{}  Goodman, A.A. 1995, in From Gas to Stars to Dust, ed. J.
Davidson, E. Erickson \& M. Haas (San Francisco: ASP), APS, vol. 73, 45
\bibitem[]{} Goodman, A.A., \& Whittet, D.C.B.\ 1995, \apjl, 455, L181
\bibitem[]{} Haslam, C.G.T. {\it et al.}\ 1982, A\&AS, 47, 1
\bibitem[]{} Hildebrand, R.H.\ 1988, QJRAS, 29, 327
\bibitem[]{} Hildebrand, R.H., Dragovan, M. 1995, \apj, 450, 663
\bibitem[]{} Hildebrand, R.H., Davidson, J.A., Dotson, J.L., Dowell, C.D.,
Novak, G. \& Vaillancourt, J.E. \ 2000, PASP, 112, 1215
\bibitem[]{}  Hildebrand, R.H., Dotson, J.L., Dowell, C.D.,
Schleuning, D.A. \& Vaillancourt, J.E. \ 1999, \apj, 516, 834 
\bibitem[]{} Jones, R.V., \& Spitzer, L., Jr.\ 1967, \apj, 147, 943
\bibitem[]{} Kamionkowski, M., Kosowski, A., \& Stebbins, A. 1997, \prl, 78,
 2058
\bibitem[]{} Kogut, A., {\it et al.}\ 1996a, \apj, 460, 1
\bibitem[]{} Kogut, A., {\it et al.}\ 1996b, \apjl, 464, L5
\bibitem[]{} Kogut, A.\ 1999 in ``Microwave Foregrounds'', 91
\bibitem[]{} Lagache, G. 2003, A\& A, accepted, astro-ph/0303335 
\bibitem[]{} Landau, L.D., \& Lifshitz, E.M. 1960, Electrodynamics
of continuous Media, Reading, MA: Addison-Wesley, p. 144
\bibitem[]{} Langston, G., Minter, A., D'Addario, L., Eberhardt, K.,
Koski, K., \& Zuber, J. 2000, AJ, 119, 2801
\bibitem[]{} Lazarian, A.\ 1994, \mnras, 268, 713
\bibitem[]{} Lazarian, A \ 1997b, \mnras, 288, 609
\bibitem[]{} Lazarian, A.\ 2000, in ``Cosmic
Evolution and Galaxy Formation'', ASP v.215, eds. Jose Franco, 
Elena Terlevich, Omar Lopez-Cruz, Itziar Aretxaga, p. 69-79, astro-ph/0003314
\bibitem[]{} Lazarian, A. 2003, JASRT, 79, 881
\bibitem[]{} Lazarian, A., \& Efroimsky, M.\ 1999, \mnras, 303, 673
\bibitem[]{} Lazarian, A., Goodman, A.A., \& Myers, P.C. 1997,
ApJ, 490, 273
\bibitem[]{} Lazarian, A., \& Draine, B.T.\ 1999, \apjl, 520, L67
\bibitem[]{} Lazarian, A., \& Draine, B.T.\ 2000, \apjl, 535, L15
\bibitem[]{} Lazarian, A., \& Roberge, W.G.\ 1997, \apj, 484, 230
\bibitem[]{} Lazarian, A., \& Prunet, S. 2002, in Astrophysical Polarized
Backgrounds,Edited by Stefano Cecchini, Stefano Cortiglioni, Robert
                     Sault, and Carla Sbarra,  AIP, Vol. 609, Melville,
p. 32
\bibitem[]{} Lazarian, A., \& Yan, H. 2002, \apj, 566, L105
\bibitem[]{} Lee, H.M., \& Draine, B.T., 1985, \apj, 290, 211
\bibitem[]{} L\'eger, A., \& Puget, J.L.\ 1984, \apjl, 278, L19
\bibitem[]{} Lesgourgues, J., Prunet, S., \& Polarski, D. \ 1999, \mnras, 303, 45
\bibitem[]{} Li, A., \& Draine, B.T. 2001, in preparation
\bibitem[]{} Martin, P.G.\ 1995, \apjl, 445, L63
\bibitem[]{} Masi, S. et al. 2001, ApJ, 553, L93
\bibitem[]{} Mathis, J.S.\ 1986, \apj, 308, 281
\bibitem[]{} Mathis, J.S., Rumpl, W., \& Nordsieck, K.H. \ 1977, \apj, 217, 425
\bibitem[]{} McCullough {\it et al.}\ 1999 ``Microwave Foregrounds'', 253
\bibitem[]{} Mukherjee, P. {\it et al.}\ 2000, astro-ph/0002305
\bibitem[]{} Omont, A.\ 1986, A\&A, 164, 159
\bibitem[]{} Padoan, P., Goodman, A., Draine, B.T., Juvela, M., Norlund, A.
\& Rognvaldsson, O.E. \ 2001, ApJ, 559, 1005
\bibitem[]{} Page, L., et al. 2003, astro-ph/0302220
\bibitem[]{} Prunet, S., \& Lazarian, A. 1999 ``Microwave Foregrounds'', 113
\bibitem[]{} Prunet, S., Sethi, S.K., Bouchet, F.R., \& Miville-Desch\^enes, M.-A. \ 1998, \aap, 339, 187
\bibitem[]{} Prunet, S., Sethi, S.K., Bouchet, F.R. \ 2000, \mnras, 314, 348 
\bibitem[]{} Purcell, E.M. 1969, On the Alignment of Interstellar Dust, 
 Physica,  41, 100
\bibitem[]{} Purcell, E.M \ 1979,  \apj, 231, 404
\bibitem[]{} Purcell, E.M., \& Spitzer, L., Jr  1971, \apj,  167, 31
\bibitem[]{} Ragot, B.R. 2002, \apj, 568, 232
\bibitem[]{}  Rao, R, Crutcher, R.M., Plambeck, R.L., Wright, M.C.H. 1998,
\apj, 502, L75
\bibitem[]{} Reach, W. T., et al. 1995, \apj, 451, 188
\bibitem[]{} Reich, P., \& Reich, W. 1988, \aaps, 74, 7
\bibitem[]{}  Roberge, W.G., \& Hanany, S. 1990, B.A.A.S., 22, 862
\bibitem[]{} Roberge, W.G., \& Lazarian, A. 1999, \mnras, 305, 615
\bibitem[]{} Savage, B.D., \& Sembach, K.R. 1996, \araa, 34, 279
\bibitem[]{} Schlegel, D. J., Finkbeiner, D. P., \& Davis M. 1998,
\apj, 500, 525 [SFD]
\bibitem[]{} Seljak, U. 1996, astro-ph/9608131
\bibitem[]{} Seljak, U., \& Zaldarriaga, M. 1997, \prl, 78, 2054
\bibitem[]{} Staggs, S.T., Gundersen, J.O., \& Church, S.E. 1999
``Microwave Foregrounds'', 299
\bibitem[]{} Tegmark et al.\ 2000, \apj, 530, 133  in ``Microwave Foregrounds'', 3
\bibitem[]{} Ward-Thompson, D., Kirk, J.M., Crutcher, R.M., Greaves, J.S.,
Holland, W.S., \& Andre, P. 2000, \apj, 537L, 135
\bibitem[]{} Weingartner, J.C., \& Draine, B.T. 2000, astro-ph/0010117
\bibitem[]{} Weingartner, J.C., \& Draine, B.T. 2001, \apj, 548, 000
\bibitem[]{} Zaldarriaga, M., Spergel, D.N., \& Seljak, U. 1997, \apj, 488, 1
\bibitem[]{} Yan, H. \& Lazarian, A. 2003, \apj, in press
\end{thebibliography}
\end{document}